\documentclass[aps,twocolumn,superscriptaddress]{revtex4}
\usepackage{amsmath,amsthm,verbatim} 
\usepackage{amssymb}
\usepackage{graphicx}
\usepackage{bbm,bm}
\usepackage[caption=false]{subfig}
\usepackage[rgb]{xcolor}
\usepackage{mathrsfs}
\usepackage{stackrel} 
\usepackage[all]{xy}
\usepackage{float} 

\usepackage{natbib}

\usepackage{enumerate,amsthm,amsmath,amssymb,color,graphicx,bbm,float,framed}

\newcommand{\beq}{\begin{equation}}
\newcommand{\eeq}{\end{equation}}

\renewcommand{\1}{\mathbbm{1}} 

\definecolor{myurlcolor}{rgb}{0,0,0.4}
\definecolor{mycitecolor}{rgb}{0,0.5,0}
\definecolor{myrefcolor}{rgb}{0.5,0,0}
\usepackage{hyperref}
\hypersetup{colorlinks,
linkcolor=myrefcolor,
citecolor=mycitecolor,
urlcolor=myurlcolor}

\makeatletter
\newtheorem*{rep@theorem}{\rep@title}
\newcommand{\newreptheorem}[2]{%
\newenvironment{rep#1}[1]{%
 \def\rep@title{#2 \ref{##1}}%
 \begin{rep@theorem}}%
 {\end{rep@theorem}}}
\makeatother

\newreptheorem{theorem}{Theorem}

\newreptheorem{lemma}{Lemma}

\usepackage{mathtools}
\usepackage[normalem]{ulem}

\begin{document}

\title{Composable security of two-way continuous-variable quantum key distribution without active symmetrization}

\author{Shouvik Ghorai}
\affiliation{LIP6, CNRS, Sorbonne Universit\'{e}, Paris, France}
\author{Eleni Diamanti}
\affiliation{LIP6, CNRS, Sorbonne Universit\'{e}, Paris, France}
\author{Anthony Leverrier}
\affiliation{Inria Paris, France}

\date{\today}

\begin{abstract}
We present a general framework encompassing a number of continuous-variable quantum key distribution protocols, including standard one-way protocols, measurement-device-independent protocols as well as some two-way protocols, or any other continuous-variable protocol involving only a Gaussian modulation of coherent states and heterodyne detection. The main interest of this framework is that the corresponding protocols are all covariant with respect to the action of the unitary group $U(n)$, implying that their security can be established thanks to a Gaussian de Finetti reduction. In particular, we give a composable security proof of two-way continuous-variable quantum key distribution against general attacks.
We also prove that no active symmetrization procedure is required for these protocols, which would otherwise make them prohibitively costly to implement.
\end{abstract}

\maketitle

Quantum key distribution (QKD) allows two distant parties, Alice and Bob with access to an untrusted quantum channel and an authenticated classical channel, to share a secret key which can later be used to encrypt classical messages. The remarkable property of QKD is that its security can be established in the information-theoretic setting, without appealing to any computational assumptions. While the first protocols relied on a discrete encoding of information and required single-photon detectors \cite{BB84,eke91}, a new generation of protocols called ``continuous-variable'' (CV) encode the information on the quadratures of the quantized electromagnetic field, allowing coherent detection to advantageously replace single-photon detection \cite{DL15}. 
There is, however, a price to pay for this simplified experimental setup and this is increased difficulty of establishing security proofs due to the fact that the finite-dimensional Hilbert space of discrete-variable QKD has to be replaced by an infinite-dimensional Fock space. Notably, the theoretical tools developed for analyzing discrete-variable protocols  \--- de Finetti theorems \cite{CKMR07,ren07,CKR09}, entropic uncertainty relations \cite{TR11}, entropy accumulation \cite{DFR16} \--- need not directly work in the CV setting.

Fortunately, some of these proof techniques have been successfully adapted to continuous variables and two one-way CVQKD protocols are now established to be secure against general attacks. These are the no-switching protocol \cite{WLB04} where Alice sends coherent states with a Gaussian modulation and Bob performs heterodyne (or dual-homodyne) detection, and the BB84-inspired protocol of Ref.~\cite{CLV01} where Alice sends squeezed states along one of the two quadratures and Bob performs homodyne detection. The security of the latter follows from a continuous-variable version of the entropic uncertainty principle \cite{FBB12} while that of the former protocol is established thanks to a recently developed Gaussian de Finetti theorem \cite{lev17,lev18}, \footnote{Alternatively, one can obtain a security proof for this specific protocol with a worse key rate in the finite-size regime by first reducing the problem to that of collective attacks using an exponential version of de Finetti theorem \cite{RC09} and then establishing the security of the protocol against collective attacks in a composable setting \cite{lev15}.}.

Establishing the security of two-way CVQKD, where Alice and Bob send quantum information back and forth through the channel, has been an outstanding goal in the field and partial progress was obtained in Refs \cite{PML08,SPS12,ZLW14,OMP15,OP16,ZLZ17,ZZS17,ZZL18}. However, to the best of our knowledge, none of these works has proven its security against general attacks in the composable setting. A notable recent result shows that for all two-way CVQKD protocols, it is sufficient to consider collective attacks \cite{ZZL18}. Unfortunately, we don't know how to analyze this restricted class of attacks in the composable setting, the main issue being the difficulty of estimating the covariance matrix of the state shared by Alice and Bob without assuming that this state is Gaussian or admits bounded higher moments for instance.

In the present paper, we explore the possibility of applying the Gaussian de Finetti reduction to CVQKD beyond one-way schemes, for instance to the measurement-device-independent (MDI) CVQKD protocol of Ref.~\cite{POS15,LZX14}, the two-way protocol with Gaussian displacements of Ref.~\cite{PML08} as well as a variant of floodlight (FL) QKD \cite{ZZD16,ZZW17,ZZS18}. What makes these protocols particularly noteworthy is that they display a symmetry with respect to the unitary group $U(n)$, where $n$ is the number of rounds of the protocols. This is a significant strengthening of the usual invariance under Alice and Bob randomly permuting their $n$ subsystems in a coordinated fashion. Recall that the standard argument for establishing the security of a protocol is to first remark that it is sufficient to consider attacks displaying the same symmetry as the protocol, and then that the usual de Finetti theorem precisely asserts that permutation-invariant states are close to independent and identically distributed (i.i.d.) states, which correspond to \textit{collective} attacks \cite{ren08}. 
For CVQKD, however, collective attacks remain nontrivial to analyze due to the infinite dimensionality of the Hilbert space. In that case, the stronger $U(n)$-symmetry allows us to exploit the Gaussian de Finetti theorem stating that it is in fact sufficient to consider a class of \textit{Gaussian} collective attacks, which turn out to be fairly simple to handle.

A potential issue raised by exploiting the invariance under $U(n)$ is whether an active symmetrization must be enforced by Alice and Bob, which would in particular require picking a Haar random matrix in $U(n)$ and make the whole protocol unpractical. Here, following the idea of Portmann \cite{por17} of dividing the QKD protocol into two parts (first identifying a min-entropy resource, and then extracting the key), we are able to show that no active symmetrization is needed for our security proof to go through, a result similar to what was known for discrete-variable protocols \cite{ren07}.

In the following, we first describe our general framework and show how it encompasses one-way as well as two-way protocols. Then we explain in which sense they are covariant with respect to the action of the unitary group $U(n)$, and how to exploit the Gaussian de Finetti  reduction to show that it is sufficient to analyze Gaussian collective attacks. Finally, we discuss in more detail the case of Gaussian two-way protocols.

{\bf A general framework for Gaussian protocols with heterodyne detection}.--- One-way CVQKD already gives rise to a zoo of different protocols depending on the states sent by Alice (coherent, squeezed or thermal), their modulation (Gaussian, discrete, along one or two quadratures) and Bob's detection (homodyne or heterodyne): see Ref.~\cite{DL15} for a recent overview of CV protocols. Two-way QKD offers even more possibilities! 
Here we will restrict our attention to the ``entanglement-based'' (EB) protocols where the honest parties prepare bipartite pure states such as two-mode squeezed vacuum states (TMSS) and exchange optical modes through an untrusted quantum channel. This is without loss of generality since any Prepare-and-Measure (PM) protocol admits an EB version with the same security \cite{GCW03}, and it is therefore sufficient to analyze the latter version even if one implements the PM scheme.

Gaussian protocols stand out among all CV protocols as the ones where Alice and Bob start out by preparing TMSS and only perform Gaussian operations and measurements (homodyne or heterodyne): indeed Gaussian attacks are then asymptotically optimal provided that the covariance matrix of the state shared by the honest parties is known, meaning that the security proof is not composable \cite{WGC06,GC06,NGA06}.
In this paper, we consider all protocols where Alice and Bob prepare TMSS, possibly perform two-mode squeezing or beamsplitter transformations (as described later) and finally measure their respective modes with heterodyne detection.

The main feature of these protocols is that they are covariant with respect to the action of the unitary group $U(n)$. Let us explain what this means. Given an $n$-mode Fock space with annihilation operators $\mathbf{a} = (a_1, \ldots, a_n)$, the unitary group acts on this space by mapping $\mathbf{a}$ to $U \mathbf{a}$, and similarly $\mathbf{a}^\dagger$ to $\overline{U} \mathbf{a}^\dagger$, where $a_i^\dagger$ is the creation operator of mode $i$, and $\overline{U}$ is the complex conjugate of $U$. A beamsplitter of transmittance $t\in [0,1]$ and a two-mode squeezing operator with gain $g\geq 1$ act on a $2$-mode Hilbert space with annihilation operators $a$ and $b$ via the respective transformations $[a, a^\dagger, b, b^\dagger]^T \to B(t) [a, a^\dagger, b, b^\dagger]^T$ and $[a, a^\dagger, b, b^\dagger]^T \to S(g) [a, a^\dagger, b, b^\dagger]^T$ 
with $B(t) = \left[\begin{smallmatrix} \sqrt{t} \mathbbm{1}_2 & -\sqrt{1-t} \mathbbm{1}_2 \\ \sqrt{1-t} \mathbbm{1}_2 & \sqrt{t} \mathbbm{1}_2 \end{smallmatrix} \right]$, $S(g)=\left[\begin{smallmatrix} \sqrt{g} \mathbbm{1}_2 & \sqrt{g-1} \sigma_x \\ \sqrt{g-1} \sigma_x & \sqrt{g} \mathbbm{1}_2 \end{smallmatrix} \right]$ and $\sigma_x = \left[\begin{smallmatrix} 0 & 1\\ 1 & 0\end{smallmatrix} \right]$ \cite{WPG12}.
Finally, heterodyne detection is nothing else than a generalized measurement where the POVM elements are given by coherent states, as follows from the resolution of the identity of an $n$-mode Fock space: $\mathbbm{1}_{\mathcal{H}}  = \frac{1}{\pi^n} \int |\bm{\alpha}\rangle \langle \bm{\alpha}| \mathrm{d} \bm{\alpha}$, with $\mathrm{d} \bm{\alpha}$ the uniform measure on $\mathbbm{C}^n$. This can be formalized as a quantum-classical map $\mathcal{M}$ defined by $\mathcal{M}(\rho) = \frac{1}{\pi^n} \int \langle \bm{\alpha} |\rho| \bm{\alpha}\rangle |\bm{\alpha}^{\mathrm{cl}} \rangle \langle \bm{\alpha}^{\mathrm{cl}}|  \mathrm{d} \bm{\alpha}$ where the superscript 'cl' means that this is a classical state encoding the value of $\bm{\alpha}$ and not a coherent state.
An important property of beamsplitters, two-mode squeezing and heterodyne detection is that they all commute with the action of the unitary group in the following sense: $[\mathcal{S}^{\otimes n}, \mathcal{U} \otimes \overline{\mathcal{U}}] = [\mathcal{B}^{\otimes n}, \mathcal{U} \otimes \mathcal{U}] =[\mathcal{M}, \mathcal{U}]  = 0$, where $\mathcal{S}, \mathcal{B}$ and $\mathcal{U}$ refer to the action of the symplectic operations $S, B$ and $U$ on their corresponding Fock space \footnote{We abuse notation slightly by considering that $\mathcal{U}$ applies to $|\bm{\alpha}^{\mathrm{cl}}\rangle$ in the obvious way, that is, by mapping $|\bm{\alpha}^{\mathrm{cl}}\rangle$ to $|U\bm{\alpha}^{\mathrm{cl}}\rangle$.}.
For this reason, protocols that start with vacuum states and where the honest parties apply two-mode squeezing (to prepare TMSS or to amplify a signal as in FL QKD), beamsplitters and perform heterodyne measurements will be covariant with respect to $U(n)$ acting as a product of the form $\mathcal{U}^{\otimes p} \otimes \overline{ \mathcal{U}}^{\otimes q}$. 

\begin{figure}[!h]
\centering
  \includegraphics[width=0.8\linewidth]{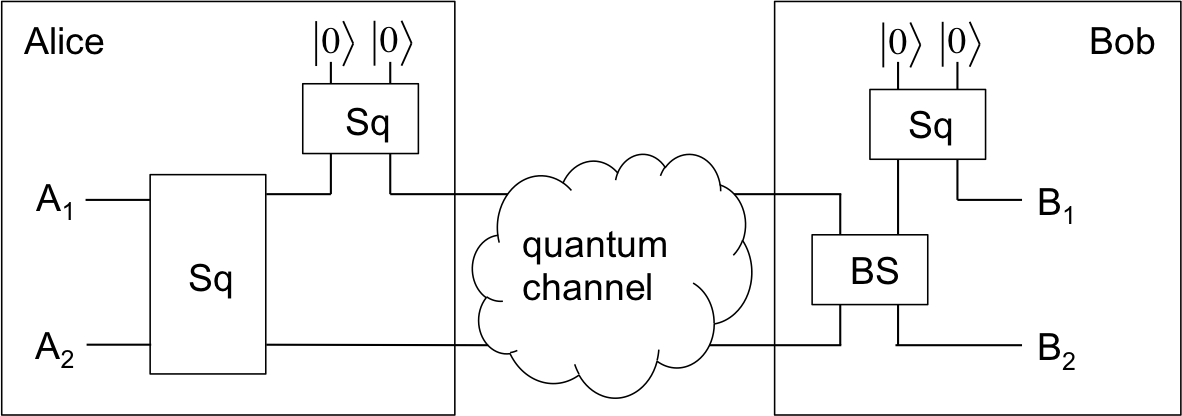}  
\caption{Optical scheme of the EB version of the two-way protocol: 'Sq' is a two-mode squeezer; 'BS' is a beamsplitter required to implement the random displacement by Bob.}  
  \label{fig:TWa}
\end{figure}

\begin{figure}[H]
\centering
  \includegraphics[width=0.8\linewidth]{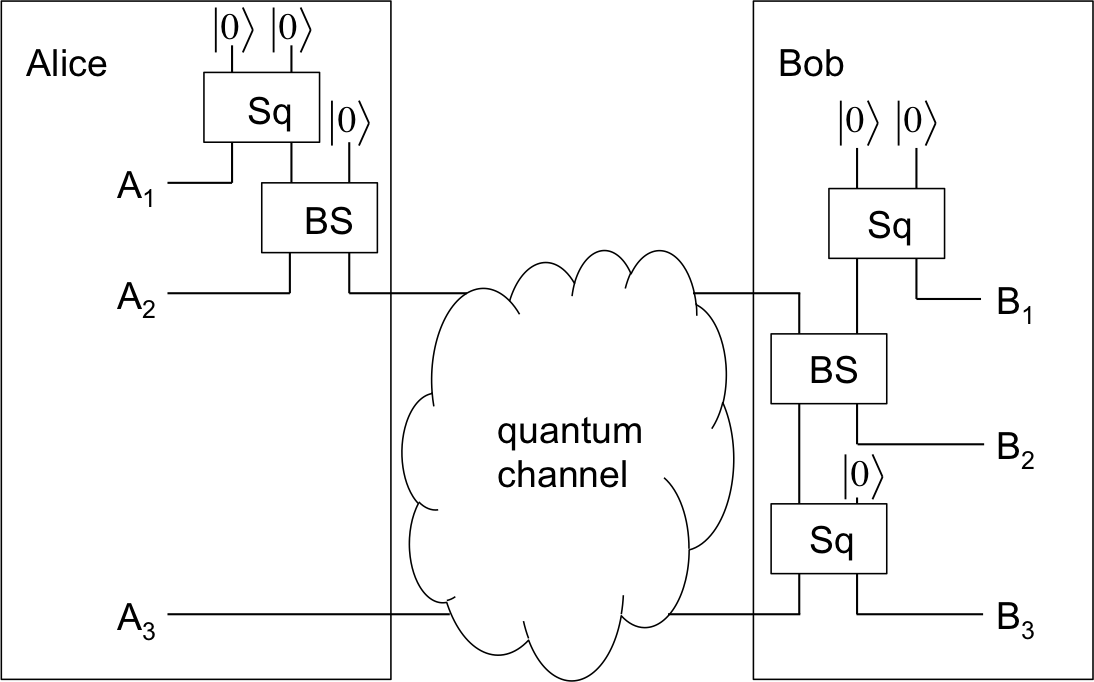}
\caption{Optical scheme of the Gaussian FL protocol}  \label{fig:FL}
\end{figure}

Examples of such CV protocols include the one-way no-switching protocol of Ref.~\cite{WLB04} and the two-way protocol with Gaussian displacements of Ref.~\cite{PML08}, whose optical schemes are respectively covariant with respect to $\mathcal{U}_A \otimes \overline{\mathcal{U}}_B$, and $\mathcal{U}_{A_1} \otimes \overline{\mathcal{U}}_{A_2} \otimes \mathcal{U}_{B_1} \otimes \overline{\mathcal{U}}_{B_2}$ (see Fig.~\ref{fig:TWa}). Our framework also allows for signal amplification thanks to two-mode squeezing as in FL QKD (see Fig.~\ref{fig:FL}).

{\bf Security proof with the Gaussian de Finetti reduction}.--- A QKD protocol typically consists of three different stages: $(i)$ state distribution, $(ii)$ measurement and parameter estimation, $(iii)$ error reconciliation and privacy amplification. 
State distribution can be modeled in two distinct ways: either by describing how Alice and Bob prepare the state and distribute it using some untrusted quantum channel (or possibly several channels), or by assuming that the state is given to them by the adversary. 
The goal of the second stage of the protocol is to obtain a \textit{min-entropy resource}, in the language of Ref.~\cite{por17}: the modes are measured with heterodyne detection, yielding classical strings, $X$ for Alice and $Y$ for Bob, one of which is then processed to give the raw key $Z= f(X)$, or $Z=f(Y)$, \textit{via} some key map $f$ \cite{ren08,CML16}. The size of the strings $X, Y$ depends on the number of modes held by the honest parties: $X, Y \in \mathbbm{C}^n$ for one-way CVQKD, while $X = (X_1, X_2), Y = (Y_1, Y_2) \in \mathbbm{C}^n \times \mathbbm{C}^n$ for two-way CVQKD, and $X, Y$ are even triplets of $n$-dimensional vectors in the case of FL protocols (see Fig.~\ref{fig:FL}). Parameter estimation then consists in a test that checks whether the correlations between $X$ and $Y$ are sufficient to imply a lower bound on the smooth min-entropy $H_{\min}^{\varepsilon}(Z|E)_{\rho_{ZE}}$, where $\varepsilon$ is the smoothing parameter and $E$ is the quantum register held by the adversary \cite{ren08}. If the test passes, the protocol continues, otherwise it aborts. As usual, one can consider without loss of generality that the whole state $\rho_{ABE}$ before measurement is pure. 
The final stage of the protocol transforms this min-entropy resource, $Z$, into a secret key using the standard techniques of QKD: error reconciliation and privacy amplification.

Assuming as usual that error reconciliation and privacy amplification are correctly implemented, it is sufficient to establish the existence of a (smooth) min-entropy resource to prove the security of the protocol \cite{por17}. In other words, we simply need to make sure that the probability that the test passes and that the min-entropy is too low is negligible, for arbitrary input states.

Let $\mathcal{M}: AB \to XY$ be the (heterodyne) measurement map and $\text{Gram}(X,Y)$ be the $m \times m$ Gram matrix (with $m=d_A+d_B$) of the vectors $X_1, \ldots, X_{d_A}, Y_1, \ldots, Y_{d_B}$ or their conjugate $\overline{X}_i$ or $\overline{Y}_j$ (depending on whether the corresponding mode is transformed according to $U$ or $\overline{U}$ through the $U(n)$-symmetry). The test $\mathcal{T}$ is then a classical function that examines $\text{Gram}(X,Y)$ (which corresponds essentially to the empirical observation of the covariance matrix) and passes if it belongs to some predefined set of acceptable covariances matrices, and fails otherwise, in which case the protocol aborts. We do not describe the set of good covariance matrices explicitly here, as this would require to take into account tedious finite-size effects. A full treatment of the security proof in the finite-size regime goes beyond the scope of this work, but the interested reader is referred to Ref.~\cite{lev15} for an example in the case of one-way CVQKD. 
The goal of the security proof is simply to show that $\mathcal{T} \circ \mathcal{M}$ realizes a min-entropy resource: the probability that the test passes and that the string $Z$ has low min-entropy with respect to Eve's register $E$ should be upper bounded by some small $\varepsilon$.

There are many subtleties involved with CVQKD, one being how to perform the parameter estimation (\textit{i.e.}, the test $\mathcal{T}$) in a composable fashion. A possibility introduced in \cite{lev15} is to postpone the parameter estimation until after error reconciliation, so that one of the parties obtains both $X$ and $Y$ and be able to compute the Gram matrix and perform the test $\mathcal{T}$. This is fine since parameter estimation and error reconciliation are classical maps that commute.

What is crucial here is that both $\mathcal{M}$ and $\mathcal{T}$ commute with the action $\mathcal{U}$ of the unitary group. For $\mathcal{T}$, this results from the choice of the set of acceptable Gram matrices, which itself is a consequence of the invariance of  the optical setup of the protocol (see Figs \ref{fig:TWa}  and \ref{fig:FL}). This implies as proven in \cite{CKR09} and \cite{lev17} that the security of the map $\mathcal{T} \circ \mathcal{M}$ (and therefore of the overall QKD protocol) can be analyzed by considering  only de Finetti states, which in our case are $SU(m,m)$ coherent states if Alice and Bob hold in total $m=d_A+d_B$ modes per round of the protocol.
As introduced in Ref.~\cite{lev18}, $SU(m,m)$ generalized coherent states are i.i.d. Gaussian states of the form $|\Lambda\rangle^{\otimes n}$ where $|\Lambda\rangle$ is a $2m$-mode Gaussian state parametrized by an $m\times m$ matrix complex $\Lambda = [\Lambda_{i,j}]$ with spectral norm $\|\Lambda\| <1$ and defined as $|\Lambda\rangle = \mathrm{det} (1-\Lambda\Lambda^\dagger)^{1/2} \exp\big( \sum_{i,j=1}^m \Lambda_{i,j} a_i^\dagger b_j^\dagger\big) |\mathrm{vacuum}\rangle$ where the creation operators of Alice and Bob's modes are denoted $a_1^\dagger, \ldots, a_m^\dagger, b_1^\dagger, \ldots, b_m^\dagger$.
For instance, $SU(1,1)$ coherent states are simply TMSS. 
In particular, the methods developed in \cite{lev17} show that $\varepsilon$-security against Gaussian collective attacks then implies $\varepsilon'$-security against general attacks, with $\varepsilon'/\varepsilon = O(n^{m^2})$. In other words, it is sufficient to show that the protocol is secure when the overall initial pure state $\rho_{ABE}$ is a mixture of such $SU(m,m)$ coherent states. 
For the protocols considered here, $m$ is in general a small constant (2 for the no-switching protocol, 4 for the two-way protocol and 6 for FL QKD), meaning that the loss in the security parameter can easily be compensated by reducing the final key length by a negligible amount. As an indication that this proof technique is rather tight, recall that the de Finetti reduction of \cite{CKR09} applied to the BB84 protocol yields a ratio $\varepsilon'/\varepsilon = O(n^{15})$ between security against general and collective attacks.

The novelty here compared to \cite{lev15,lev17} is that we restrict our analysis to $\mathcal{T} \circ \mathcal{M}$ instead of the whole QKD protocol. While the latter doesn't commute with the unitary group (because of error reconciliation for instance) and one had to symmetrize the state in order to enforce the correct invariance, the min-entropy resource part of the protocol commutes with the action of $U(n)$ and the work of Portman allows us to infer the security of the QKD protocol \cite{por17}. For this reason, the proof holds without any need for actively symmetrizing the state, even at the classical level. 
This settles an open question from Refs \cite{lev15,lev17}.

{\bf About MDI CVQKD}.--- At first sight, MDI CVQKD doesn't quite fit our framework since it involves a third node, controlled by Charlie, performing a Bell measurement consisting of \textit{homodyning} two modes. More precisely, the idea is that both Alice and Bob prepare a TMSS, keep one mode each and send their other mode to Charlie who performs entanglement swapping, publicly announcing the results of his Bell measurement, and allowing Alice and Bob to conditionally displace their remaining mode in order to create some correlations \cite{POS15}. 
In this scenario, one could \textit{a priori} consider that there are four optical modes: one for Alice, one for Bob and two measured by Charlie, hinting that one should appeal to a proof technique similar to that of two-way CVQKD. This is for instance an approach followed in \cite{LOP18} where it was realized that this scheme has the advantage of not requiring much public communication for parameter estimation. However, this description doesn't seem compatible with the Gaussian de Finetti reduction since the homodyne detection performed by Charlie breaks the invariance of the protocol under the group $U(n)$. An alternative approach is to view this scheme as a special case of one-way CVQKD by treating Charlie's communication as part of the state distribution: once Alice and Bob's displacements have been performed, the two honest parties are left with a bipartite two-mode quantum state. This is the same situation as after state distribution in the EB version of the no-switching protocol \cite{WLB04}.  
In this sense, while MDI CVQKD is implemented similarly as a two-way CVQKD protocol, its security can be analyzed as a one-way CVQKD, with a Gaussian de Finetti reduction involving $SU(2,2)$ coherent states. In particular, the reduction from Ref.~\cite{lev17} together with the security proof of Ref.~\cite{lev15} establish the security of MDI CVQKD against general attacks (see also \cite{ZZZ17,LOP18b}).

{\bf An example: two-way Gaussian CVQKD with heterodyne detection}.--- In the PM version, Alice sends coherent states with a Gaussian modulation to Bob, who performs a random Gaussian displacement to the mode he receives and sends it back through the quantum channel. Alice measures the output mode with heterodyne detection, and computes a weighted sum of this result and the value of her initial coherent state. This serves as the raw key. The weights in the sum as well as the variances of the Gaussian distribution should be optimized to yield the maximum key rate. Note that this protocol differs a little bit from the one of Ref.~\cite{PML08} in that here Bob always performs a displacement, that the weights of the sum are optimized and that Bob also exploits the second output of his beamsplitter to guess the raw key.
The setup of the corresponding EB version is depicted on Fig.~\ref{fig:TWa}.

If the quantum channels are covariant with respect to $U(n)$, as expected for instance in the case of a passive adversary and a bosonic phase-insensitive channel, then the covariance matrix of the state $\rho_{A_1A_2B_1 B2}$ is invariant under $\mathcal{U}_{A_1} \! \otimes \overline{\mathcal{U}}_{A_2} \! \otimes  \! \ \mathcal{U}_{B_1} \otimes \! \ \overline{\mathcal{U}}_{B_2}$, as detailed in Appendix \ref{sec:symm}. For this protocol, the set of acceptable covariance matrices will therefore satisfy the same symmetry. This implies that the Gaussian de Finetti reduction can be applied to provide a composable security proof against general attacks. In particular, this means that Gaussian attacks, described by $SU(4,4)$ generalized coherent states, are asymptotically optimal and that the asymptotic key rate can be computed with standard techniques \cite{WPG12}: see Appendix \ref{sec:rate} for details. 

{\bf About Floodlight (FL) QKD}.---FL QKD offers the perspective of much higher key rates than traditional QKD at metropolitan range by exploiting multimode encoding and signal amplification techniques to compensate for the extra losses present in two-way QKD compared to one-way QKD. However, establishing the security of this scheme has proven challenging. Here, we note that variants of FL QKD fit our framework, provided that the encoding is done through a random Gaussian displacement by Bob, instead of applying a random phase to the signal. Indeed, such a (binary) phase-shift breaks the $U(n)$-symmetry of the protocol. A description of such a Gaussian protocol appears on Fig.~\ref{fig:FL}. There are two additional elements compared to two-way CVQKD: Alice attenuates her signal before sending it through the quantum channel with a beamsplitter, and Bob is allowed to amplify his signal with a two-mode squeezer before sending it back to Alice. In total, both Alice and Bob hold three optical modes for each of the $n$ instances, and the protocol is invariant under the action of the unitary group $U(n)$ acting as $\mathcal{U}_{A_1} \! \otimes  \overline{\mathcal{U}}_{A_2}\!  \otimes \overline{\mathcal{U}}_{A_3}\!  \otimes \mathcal{U}_{B_1}\!  \otimes \overline{\mathcal{U}}_{B_2} \! \otimes {\mathcal{U}}_{B_3}$.
In particular, this implies that one can use the Gaussian de Finetti reduction to reduce the security proof to considering Gaussian attacks, parametrized by $SU(6,6)$ generalized coherent states.

{\bf Conclusion}.--- In this work, we considered a large class of CVQKD protocols which are invariant with respect to the unitary group $U(n)$ and showed that it is sufficient to establish their security against Gaussian collective attacks. This extends the results of Ref.~\cite{lev17} to two-way protocols which are known to display improved tolerance to noise compared to the no-switching protocol, and provides the first composable security proof for two-way CVQKD protocols against general attacks. 
Moreover, by exploiting the modularity of the QKD protocols as introduced by Portmann, we proved that active symmetrization of the data is not needed to apply the de Finetti reduction and to obtain security.



\acknowledgements{We acknowledge funding from European Union's Horizon 2020 research and innovation programme under the Marie Sk\l{}odowska-Curie grant agreements No 675662 and  No 820466 (CiViQ), and French National Research Agency (ANR) project quBIC.}

\newpage

\begin{widetext}

\appendix

In this appendix, we first study in detail in Section \ref{sec:symm} why the two-way and the FloodLight CV QKD protocols respect the symmetry of the unitary group $U(n)$. Then for completeness, we provide in Section \ref{sec:rate} the explicit calculations of the asymptotic secret key rate of our version of the two-way protocol.

\section{Symmetry of the two-way and floodlight protocols with respect to $U(n)$.}
\label{sec:symm}

\subsection{Two-way protocol}
\label{subsec:TW}

In order to show that the first part of the protocol (which implements a min-entropy resource) is covariant with respect to the action of $U(n)$, we need to show that both the optical scheme (in Alice and Bob's labs) and the parameter estimation procedure satisfy this symmetry.
We have not completely specified the parameter estimation procedure so far, except to say that it consists in estimating the Gram matrix of the measurement outcomes of Alice and Bob, and accepting if the estimate is within some predefined acceptance region, or aborting otherwise. The protocol (and in particular, the acceptance region) is in general designed in order to perform well when the adversary is passive: here, it means that the quantum channel is expected to be a covariant bosonic thermal channel which is a good model for fiber-based quantum communication.
Under the assumption that both quantum channels are covariant, it is straightforward to verify graphically (as depicted on Fig.~\ref{fig:TWU}) that the quantum state held in registers $A_1, A_2, B_1, B_2$ by Alice and Bob before they perform their heterodyne measurement is invariant under the unitary transformation $ \mathcal{U} \otimes \overline{\mathcal{U}} \otimes  \mathcal{U} \otimes \overline{\mathcal{U}}$. 
For this reason, it is natural to choose an accepting region for the parameter estimation test that also satisfies this symmetry.

Now the parameter estimation procedure is therefore covariant with respect to the unitary group by construction, and the optical part of the protocol is also covariant as can be checked by exploiting the commutation relations 
\begin{align*}
[\mathcal{S}^{\otimes n}, \mathcal{U} \otimes \overline{\mathcal{U}}] = [\mathcal{B}^{\otimes n}, \mathcal{U} \otimes \mathcal{U}] =[\mathcal{M}, \mathcal{U}]  = 0.
\end{align*}
This means that the first part of the protocol, that implements the min-entropy resource, is indeed covariant with respect to the action of the unitary group $U(n)$, and that one can apply the de Finetti reduction in order to prove the security of the protocol against general attacks.

\begin{figure}[!h]
  \begin{center}
    \subfloat[Unitaries applied to vacuum]{
      \includegraphics[width=0.45\textwidth]{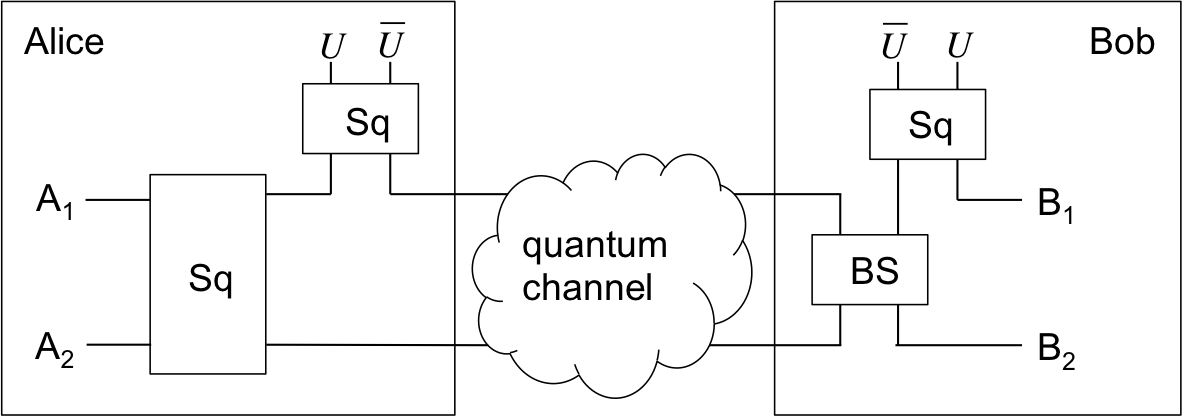}
      \label{sub:renonc}
                         }
                         \hspace{1cm}
    \subfloat[Covariance with the initial squeezers]{
      \includegraphics[width=0.45\textwidth]{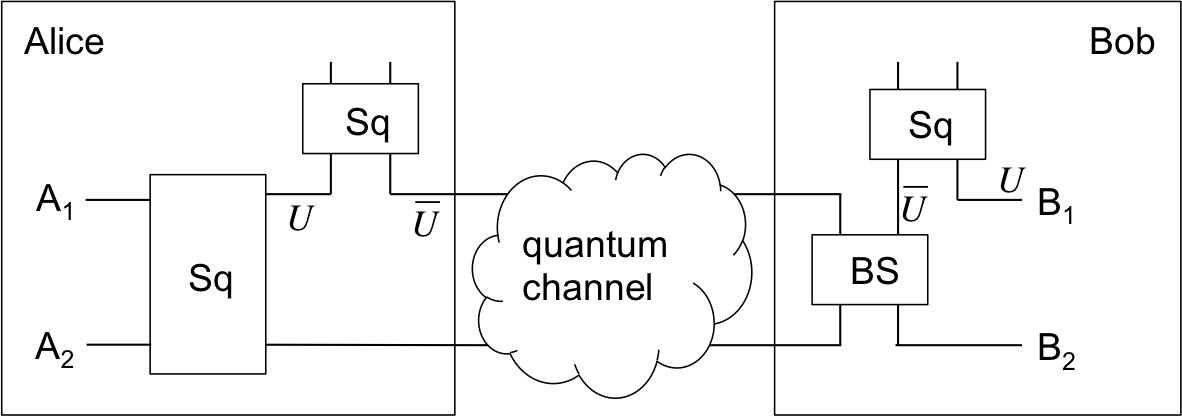}
      \label{sub:popul}
                         }\\
        \subfloat[Covariance with the forward channel]{
      \includegraphics[width=0.45\textwidth]{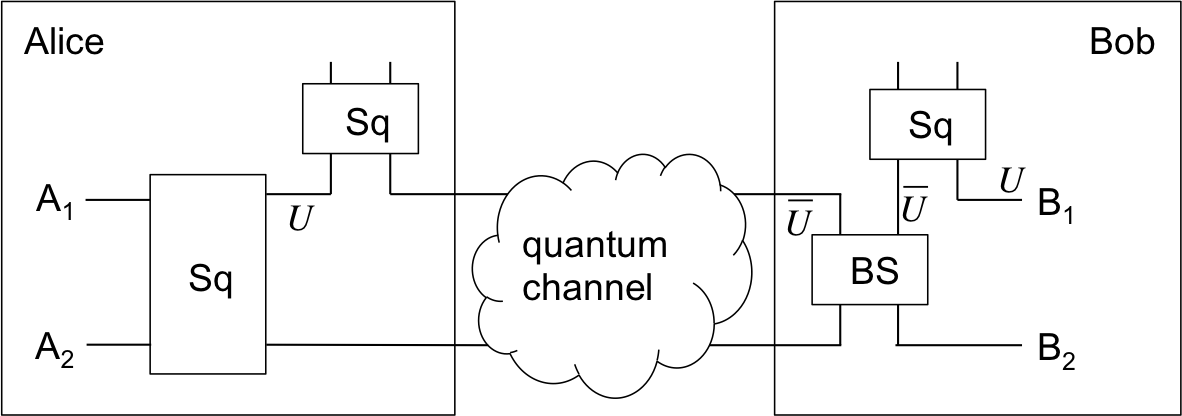}
      \label{sub:renonc}
                         }
                         \hspace{1cm}
    \subfloat[Covariance with Bob's beamsplitter]{
      \includegraphics[width=0.45\textwidth]{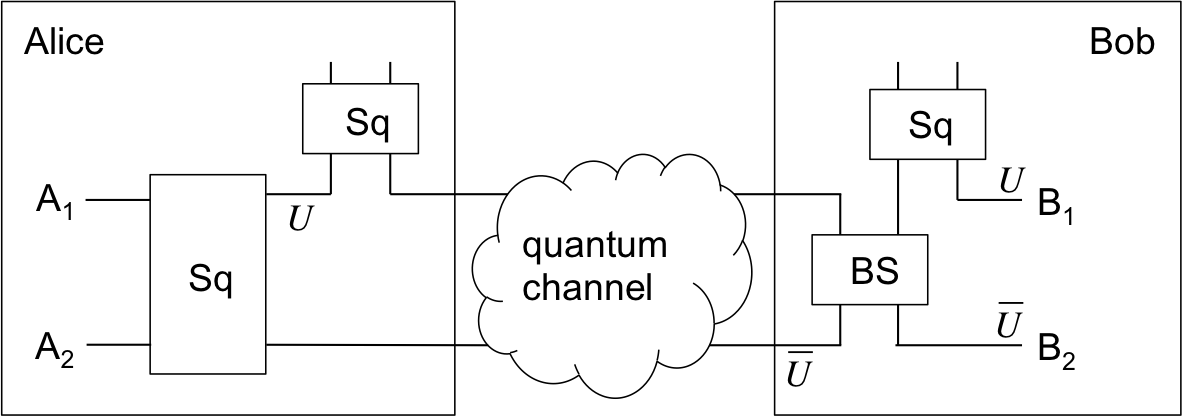}
      \label{sub:popul}
                         }\\
                           \subfloat[Covariance with the backward channel]{
      \includegraphics[width=0.45\textwidth]{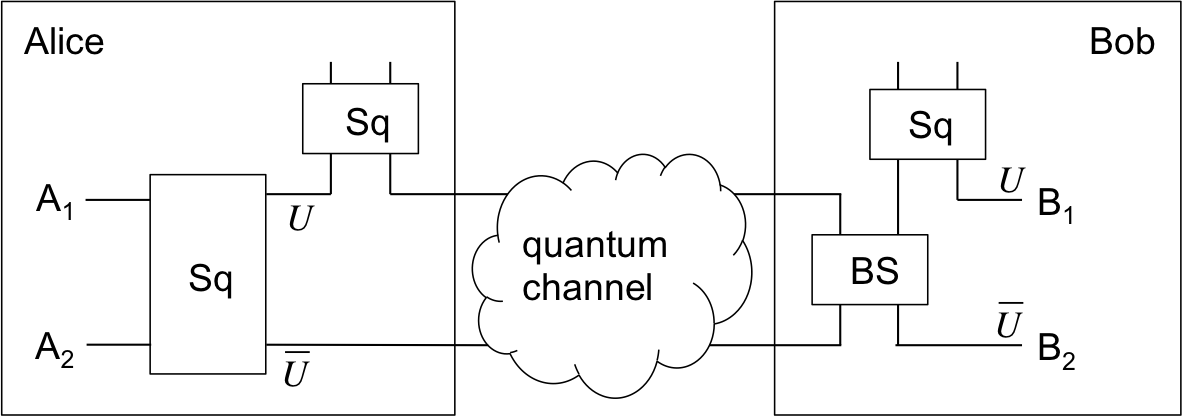}
      \label{sub:renonc}
                         }
                         \hspace{1cm}
    \subfloat[Commutation with Alice's second squeezer]{
      \includegraphics[width=0.45\textwidth]{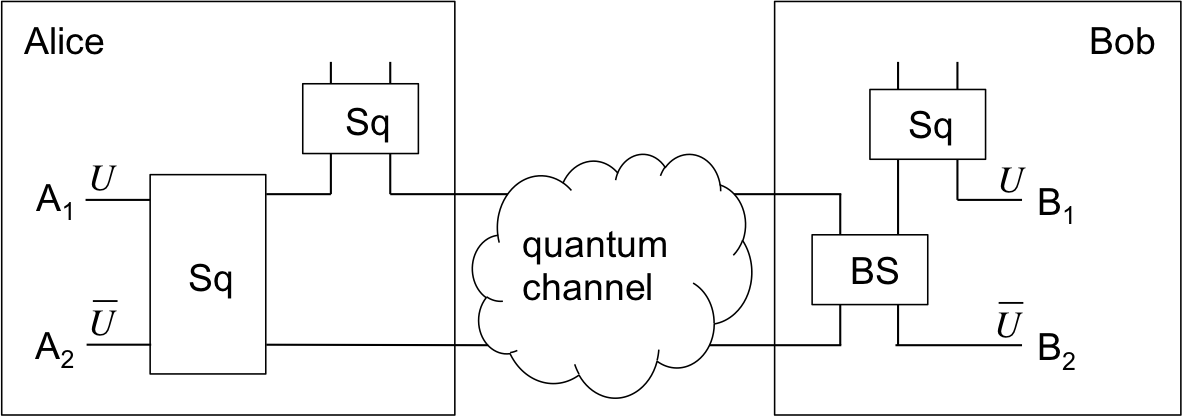}
      \label{sub:popul}
                         }\\

    \caption{Propagation of the unitaries through the circuit of the two-way protocol. `Sq' and `BS' stand respectively for two-mode squeezer and beamsplitter. One starts by applying unitaries $U$ or $\bar{U}$ to the input of the protocol and propagates these operators through the setup. Since the input of the setup is the vacuum and therefore invariant under the action of $U$ or $\bar{U}$, we infer that the protocol is invariant if Alice and Bob process their modes $A_1, A_2, B_1, B_2$ by $U \otimes \bar{U} \otimes {U} \otimes \bar{U}$.}  \label{fig:TWU}
  \end{center}
\end{figure}

\subsection{Floodlight protocol}
\label{subsec:FL}

\begin{figure}[H]
  \begin{center}
    \subfloat[Unitaries applied to vacuum]{
      \includegraphics[width=0.4\textwidth]{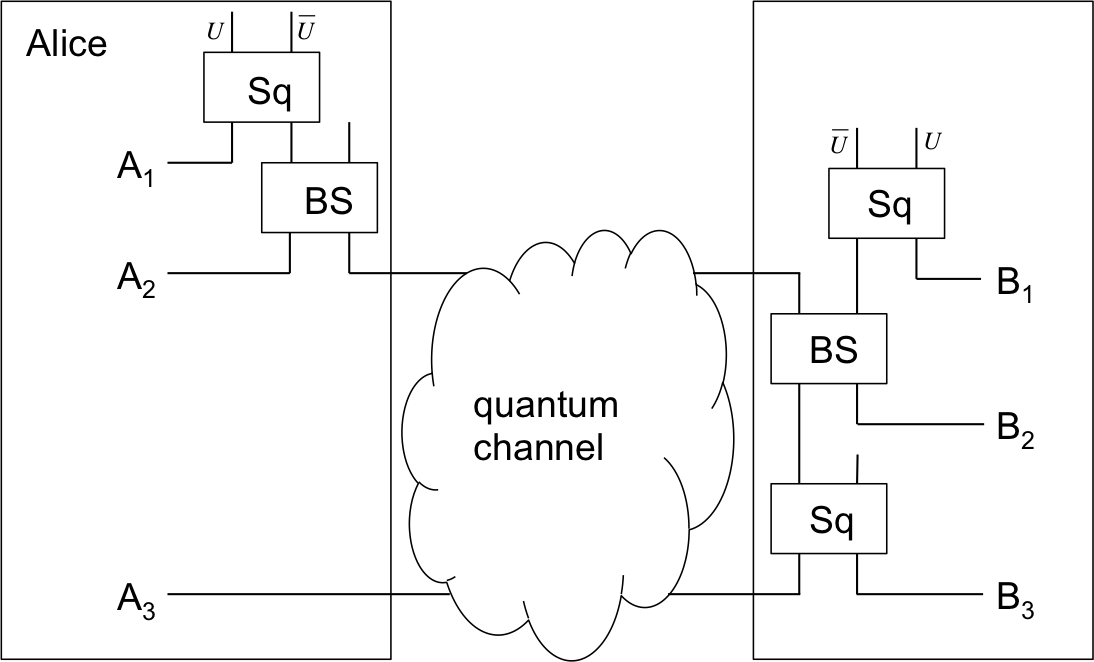}
      \label{sub:renonc}
                         }
                         \hspace{1cm}
    \subfloat[Covariance with the initial squeezers]{
      \includegraphics[width=0.4\textwidth]{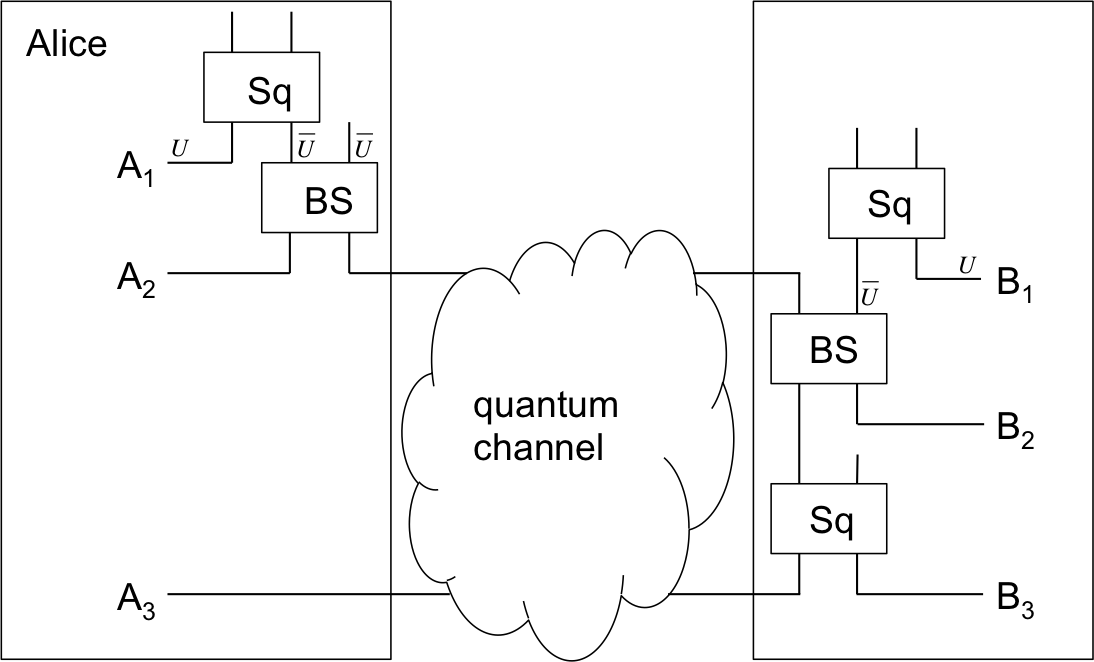}
      \label{sub:popul}
                         }\\
    \subfloat[Covariance with Alice's beamsplitter]{
      \includegraphics[width=0.4\textwidth]{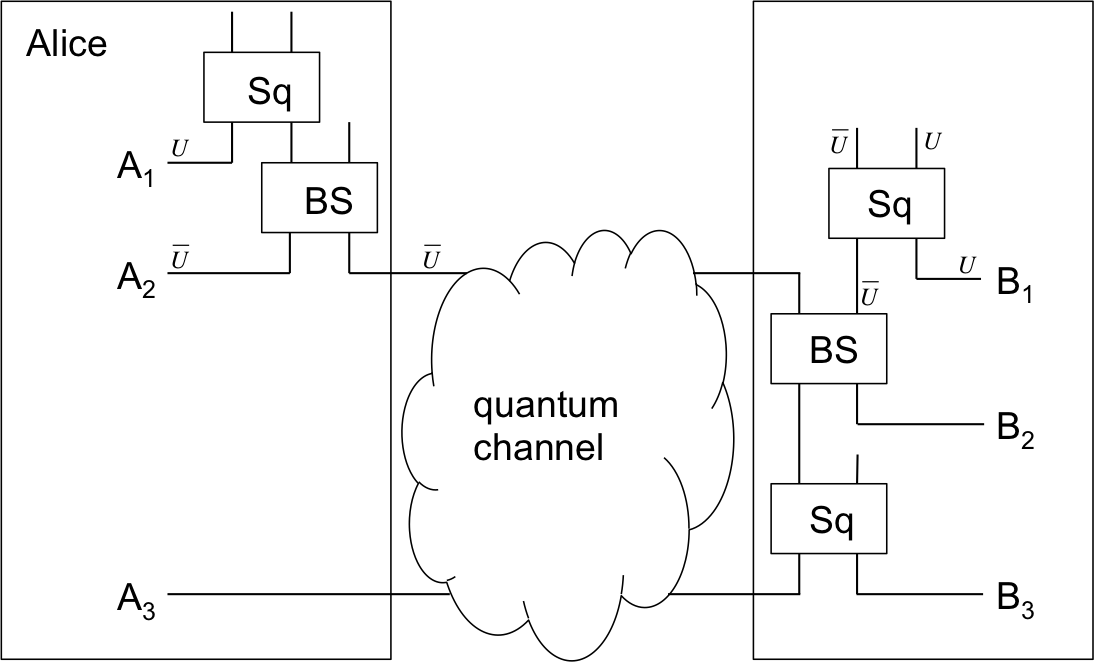}
      \label{sub:renonc}
                         }
                         \hspace{1cm}
    \subfloat[Covariance with the forward channel]{
      \includegraphics[width=0.4\textwidth]{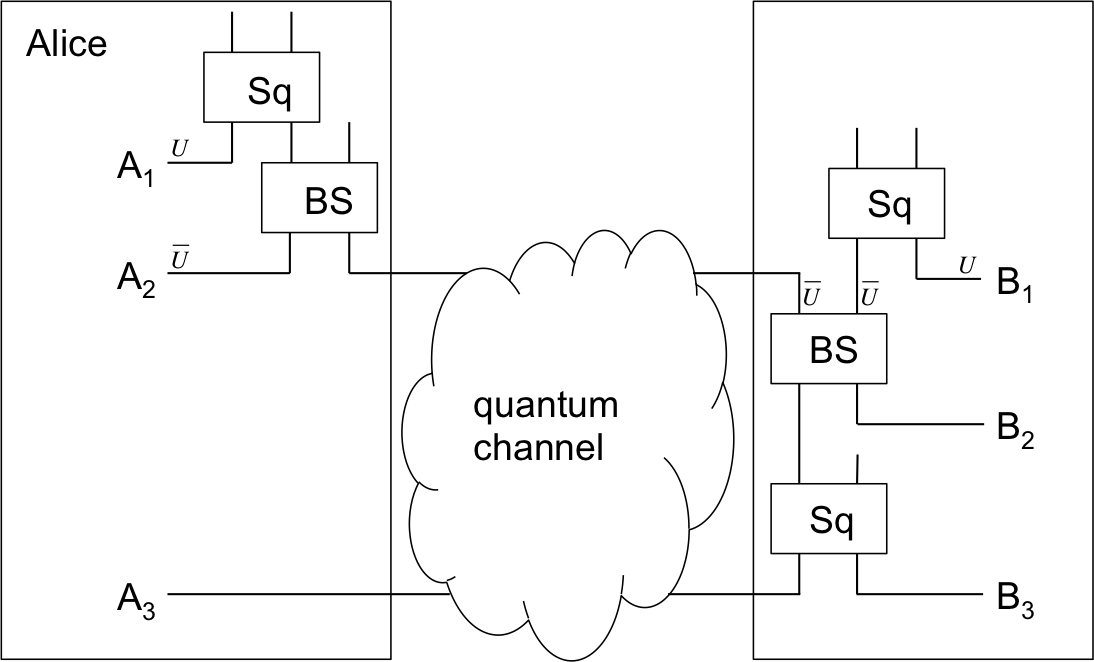}
      \label{sub:popul}
                         }\\
                           \subfloat[Covariance with Bob's beamsplitter]{
      \includegraphics[width=0.4\textwidth]{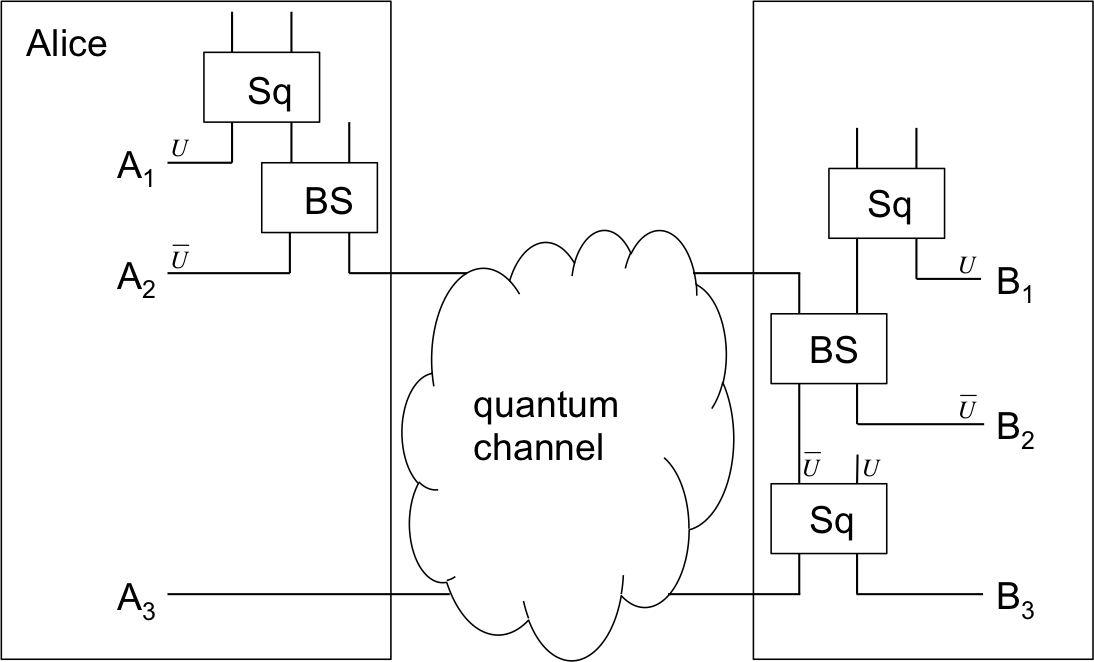}
      \label{sub:renonc}
                         }
                         \hspace{1cm}
    \subfloat[Covariance with Bob's second squeezer]{
      \includegraphics[width=0.4\textwidth]{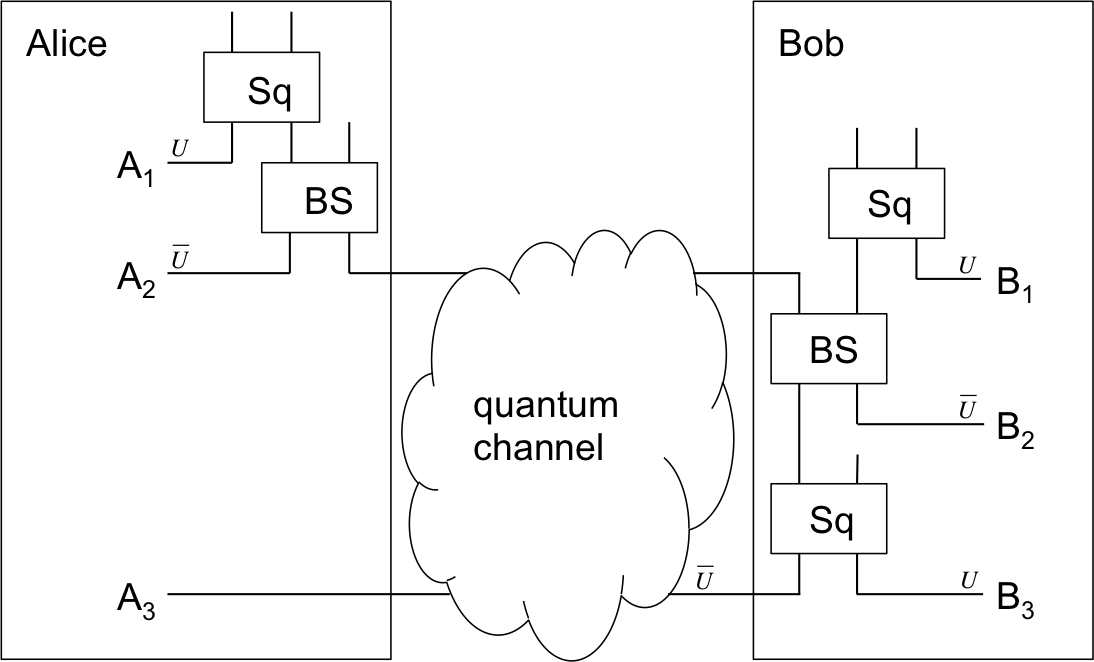}
      \label{sub:popul}
                         }\\
                           \subfloat[Covariance with the backward channel]{
      \includegraphics[width=0.4\textwidth]{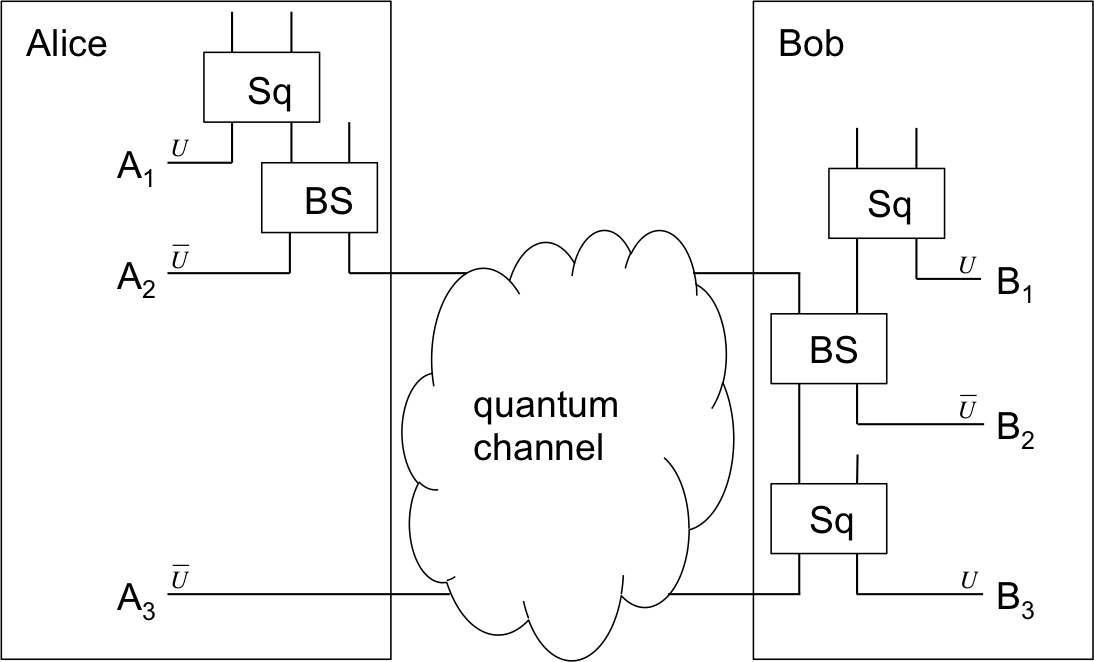}
      \label{sub:renonc}
                         }
                        
    \caption{Propagation of the unitaries through the circuit of the FL protocol.`Sq' and `BS' stand respectively for two-mode squeezer and beamsplitter. One starts by applying unitaries $U$ or $\bar{U}$ to the input of the protocol and propagates these operators through the setup. Since the input of the setup is the vacuum and therefore invariant under the action of $U$ or $\bar{U}$, we infer that the protocol is invariant if Alice and Bob process their modes $A_1, A_2, A_3, B_1, B_2, B_3$ by $U \otimes \bar{U} \otimes \bar{U} \otimes U \otimes \bar{U} \otimes U$.}  \label{fig:FLU}
  \end{center}
\end{figure}

The analysis is similar to the case of two-way CV QKD: one simply follows the propagation of unitaries through the quantum circuit, as depicted on Fig.~\ref{fig:FLU}.

\section{Secret key rate of the two-way protocol}
\label{sec:rate}

We have seen in the main text how to analyze the security of some Gaussian CV QKD protocols. Such an analysis doesn't require any assumption about the untrusted quantum channels controlled by the adversary. In order to assess the performance of QKD protocols, however, it is necessary to provide a (realistic) model of the expected quantum channels. In the context of continuous-variable communication, a standard approach is to model the quantum channels via thermal bosonic channels. Such channels are parametrized by two quantities corresponding to loss and noise: $\tau \in [0,1]$ is the transmittance of the channel and $\xi \geq 0$ is the so-called excess noise. (An alternate parametrization of the channel that is sometimes used in the literature is via transmittance and thermal noise, but we will prefer the excess noise which is the quantity more directly related to the quality of the implementation.)  With these notations, preparing an initial TMSS with variance $V \geq 1$ and covariance matrix $ \left[\begin{smallmatrix} V \mathbbm{1}_2 & \sqrt{V^2-1} \sigma_z \\ \sqrt{V^2-1} \sigma_z & V \mathbbm{1}_2\end{smallmatrix} \right]$ and sending the second mode through the channel of parameters $(\tau, \xi)$ yields a Gaussian state with covariance matrix $ \left[\begin{smallmatrix} V \mathbbm{1}_2 & \sqrt{\tau}\sqrt{V^2-1} \sigma_z \\ \sqrt{\tau}\sqrt{V^2-1} \sigma_z & (\tau(V-1+\xi) + 1)  \mathbbm{1}_2\end{smallmatrix} \right]$, where $\sigma_z = \left[\begin{smallmatrix} 1 & 0 \\ 0 & -1\end{smallmatrix} \right]$.

In this section, we review how to estimate the (asymptotic) secret key rate of this protocol. Similar analyses can be found in Refs~\cite{PML08,WOP14,OMP15,OP16} for instance.

We use the notations depicted on Fig.~\ref{fig:TWc}. Alice and Bob each start with a two-mode squeezed vacuum state, of respective variances $V_A$ and $V_B$. Alice sends one mode ($A_1'$) through the quantum channel and Bob mixes one of his modes ($B_1'$) together with the output of the quantum channel ($C_1$) on a beamsplitter of transmittance $T$. Bob then sends back one of the output modes back through the quantum channel. Alice finally performs a two-mode squeezing operation with squeezing $g$ on her two modes.
All the mode are measured with heterodyne detection, and we choose the outcome corresponding to mode $A_2$ to be the raw key. 
The values of the three squeezing operations and of the beamsplitter are optimized so as to maximize the secret key rate.

\begin{figure}[H]
\centering
  \includegraphics[width=0.7\linewidth]{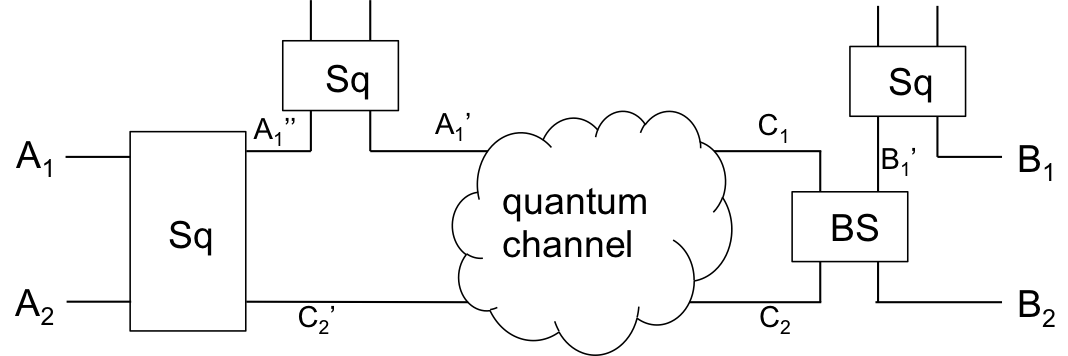}
\caption{Description of the EB version of the two-way protocol. Two pairs of two-mode squeezed vacuum states are initially prepared by Alice and Bob in modes $(A_1', A_1'')$ and $(B_1, B_1')$, respectively. Mode $A_1'$ is sent through the quantum channel. The output mode $C_1$ is combined with $B_1'$ in a beamsplitter: this effectively implements a Gaussian displacement of mode $C_1$. One output mode of the beamsplitter, $C_2$, is then sent back through the quantum channel and finally combined with $A_1''$ with a two-mode squeezer. One of the two output modes, $A_2$ is then measured with heterodyne detection and the measurement result, $X_2$, serves as the raw key. In the Prepare and Measure version of the protocol, Alice and Bob would simply prepare $A_1'$ and $B_1'$ in a coherent state with a Gaussian modulation, Bob would apply the displacement operation (via the beamsplitter) and send back the output mode $C_2$ to Alice who would measure it with heterodyne detection. None of the three two-mode squeezers are needed in the PM version as they can always be simulated classically.}  \label{fig:TWc}
\end{figure}

\subsection*{Covariance matrices}

For the purpose of assessing the performance of the protocol, it is useful to compute the key rate in realistic conditions. Here, we choose to model the quantum channels by thermal channels with transmittance $\tau$ and excess noise $\xi$.

Let us follow the notations of Fig.~\ref{fig:TWc} and compute the various covariance matrices, the one of interest being $\Gamma_{A_1 A_2 B_1 B_2}$.
We also use the following conventions: 
\begin{align*}
V_A = v+1, \quad z = \sqrt{v^2+2v},\quad 
V_B = v'+1, \quad z' = \sqrt{v'^2+2v'}.
\end{align*}

Moreover, all the matrices are block-matrices. We use boldface to indicate that a $2\times 2$ matrix is proportional to $\sigma_z$. Otherwise, the block is simply proportional to $\mathbbm{1}_2$.

The initial covariance matrix is:

\begin{align*}
\Gamma_{A_1'' A_1' B_1 B_1'} = \1_4 + \begin{bmatrix}
v&  \bm{z} &  & \\
\bm{z} & v  & & \\
& & v' & \bm{z'} \\
& & \bm{z'} & v'
\end{bmatrix},
\end{align*}
where we omit the zeroes in the matrices.

Mode $A_1'$ goes through the quantum channel $(\tau, \xi)$, yielding:
\begin{align*}
\Gamma_{A_1'' C_1 B_1 B_1'} = \1_4 + \begin{bmatrix}
v&  \bm{\sqrt{\tau} z} &  & \\
\bm{\sqrt{\tau} z} & \tau(v+\xi)  & & \\
& & v' & \bm{z'} \\
& & \bm{z'} & v'
\end{bmatrix}.
\end{align*}

The beamsplitter interaction gives
\begin{align*}
\Gamma_{A_1'' C_2 B_2 B_1} &= \left(\mathbbm{1}_{2} \oplus \begin{bmatrix} \sqrt{T} \mathbbm{1}_{2} &   \sqrt{1-T} \mathbbm{1}_{2} \\- \sqrt{1-T} \mathbbm{1}_{2} & \sqrt{T} \mathbbm{1}_{2} \end{bmatrix} \oplus \mathbbm{1}_2 \right)  \Gamma_{A_1 C_1 B_1' B_1} \left(\mathbbm{1}_{2} \oplus \begin{bmatrix} \sqrt{T} \mathbbm{1}_{2} &  - \sqrt{1-T} \mathbbm{1}_{2} \\ \sqrt{1-T} \mathbbm{1}_{2} & \sqrt{T} \mathbbm{1}_{2} \end{bmatrix} \oplus \mathbbm{1}_2 \right)\\
&= \1_4 + \begin{bmatrix}
v & \bm{\sqrt{T\tau} z} & \bm{-\sqrt{(1-T)\tau} z} & \\
* & T\tau (v+\xi) +(1-T) v' & -\sqrt{T(1-T)} (\tau(v+\xi) - v')& \bm{\sqrt{(1-T)} z'}\\
* & * & (1-T)\tau(v+\xi) + T v' & \bm{\sqrt{T} z' }\\
 & *& * & v' \\
\end{bmatrix}.
\end{align*}
Since the matrices are symmetric, we simply write $*$ in the bottom left matrix to improve the readability. 

Now mode $C_2$ goes back through the quantum channel $(\tau, \xi)$. One obtains:
\begin{align*}
\Gamma_{A_1'' C_2' B_2 B_1} &=  \1_4 + \begin{bmatrix}
v & \bm{\sqrt{T}\tau z} & \bm{-\sqrt{(1-T)\tau} z} & \\
* & T\tau^2 (v+\xi) +(1-T) \tau v' + \tau \xi & -\sqrt{T(1-T)\tau} (\tau(v+\xi) - v')& \bm{\sqrt{(1-T)\tau} z'}\\
* & * & (1-T)\tau(v+\xi) + T v' & \bm{\sqrt{T} z' }\\
 & *& * & v' \\
\end{bmatrix}\\
&= \1_4 +  \begin{bmatrix}
v & \bm{z_1} & \bm{z_2} & \\
* & v_1 & z_{12}& \bm{z_1'}\\
* & * &v_2 & \bm{z_2' }\\
 & *& * & v' \\
\end{bmatrix} = 
 \begin{bmatrix}
V & \bm{z_1} & \bm{z_2} & \\
* & V_1 & z_{12}& \bm{z_1'}\\
* & * &V_2 & \bm{z_2' }\\
 & *& * & V' \\
\end{bmatrix}
\end{align*}
with 
\begin{align*}
v_1 &:= T\tau^2 (v+\xi) +(1-T) \tau v' + \tau \xi\\
v_2 & :=  (1-T)\tau(v+\xi) + T v'\\
z_1 & := \sqrt{T}\tau z\\
z_2 & := -\sqrt{(1-T)\tau} z\\
z_{12} & := -\sqrt{T(1-T)\tau} (\tau(v+\xi) - v') \\
z_1' & := \sqrt{(1-T)\tau} z' \\
z_2' & := \sqrt{T}z' \\
V_i &:= 1+v_i
\end{align*}

Finally, Alice applies two-mode squeezing to her two modes, in order to form the raw key. This can be interpreted as performing a weighted combination of her two modes. The value of the squeezing is optimized so as to maximize the expected secret key.  
\begin{align*}
\Gamma_{A_1 A_2 B_2 B_1}  = \left(  \begin{bmatrix} \sqrt{g} & \bm{-\sqrt{g-1}} \\ \bm{-\sqrt{g-1}} & \sqrt{g} \end{bmatrix} \oplus \1_4\right) \Gamma_{A_1'' C_2' B_2'' B_1''}  \left(\begin{bmatrix} \sqrt{g} & \bm{-\sqrt{g-1}} \\ \bm{-\sqrt{g-1}} & \sqrt{g} \end{bmatrix} \oplus \1_4 \right)
\end{align*}
with $g \geq 1$.

This gives:
\begin{align*}
&\Gamma_{A_1 A_2 B_2 B_1}=\\
 &
 \begin{bmatrix}
g V + (g-1) V_1 + 2\sqrt{g(g-1)} z_1 & \bm{-\sqrt{g(g-1)} (V+V_1) + (2g-1)z_1} & \bm{\sqrt{g} z_2 - \sqrt{g-1} z_{12}} &- \sqrt{g-1} z_1'\\
* & (g-1) V + g V_1 - 2\sqrt{g(g-1)} z_1 &- \sqrt{g-1} z_2 + \sqrt{g} z_{12}& \bm{\sqrt{g} z_1'}\\
* & * &V_2 & \bm{z_2' }\\
 & *& * & V' \\
\end{bmatrix}
\end{align*}

The raw key is defined to be $X_2$, the measurement outcome obtained when performing heterodyne detection on mode $A_2$.

In order to compute the secret key rate, it will also be useful to compute the covariance matrix $\Gamma_{A_1 B_2 B_1 |X_2}$ of the state $A_1 B_2 B_1$ conditioned on the measurement result $X_2$. 
Given a general covariance matrix of the form $\Gamma = \left[ \begin{smallmatrix} A & C \\ C^T & B\end{smallmatrix}\right]$, the covariance matrix of the state conditioned on the measurement result of heterodyning the modes corresponding to block $A$ is \cite{ESP02}
\begin{align*}
\Gamma_{A_1 B_2 B_1 |X_2}= B - C^T (A + \1_2)^{-1} C,
\end{align*}
with the following submatrices in our case where Alice measures the second mode with heterodyne detection:
\begin{align*}
A &:=  ( (g-1) V + g V_1 - 2\sqrt{g(g-1)} z_1-) \1_2,\\
B &:=  \begin{bmatrix} g V + (g-1) V_1 + 2\sqrt{g(g-1)} z_1 & \bm{\sqrt{g} z_2 - \sqrt{g-1} z_{12}} &- \sqrt{g-1} z_1'\\
* &V_2 & \bm{z_2' }\\
* & * & V' \\
  \end{bmatrix},\\
C &:=  \begin{bmatrix}  \bm{-\sqrt{g(g-1)} (V+V_1) + (2g-1)z_1} & - \sqrt{g-1} z_2 + \sqrt{g} z_{12}& \bm{\sqrt{g} z_1'}
  \end{bmatrix}.
 \end{align*}

\subsection*{Secret key rate}

With the above covariance matrices at hand, we are now ready to compute the asymptotic key rate of the two-way protocol for a typical thermal bosonic channel of transmittance $\tau$ and excess noise $\xi$.

Exploiting the Gaussian de Finetti reduction as explained in the main text, the asymptotic key rate is given by the Devetak-Winter formula \cite{DW05}:
It reads:
\begin{align}
K &= \frac{1}{2} \left( \beta I (X_2; (Y_1, Y_2)) - \chi(X_2;E) \right),
\end{align}
where we choose $X_2$, the measurement outcome of heterodyning mode $A_2$, to be the raw key. Here, $\beta \leq 1$ is the so-called ``reconciliation efficiency''. The factor $1/2$ reflects the fact that we compute a key rate per channel use, and that two channels are used in two-way CV QKD.

Let $\Gamma_{X_1 X_2 Y_1 Y_2}$ denote the covariance matrix of the measurement outcomes. It if given by:
\begin{align*}
\Gamma_{X_1 X_2 Y_1 Y_2} = \frac{1}{2} (\Gamma_{A_1 A_2 B_1 B_2} + \1_8).
\end{align*}

A subtlety of the EB protocol is that the parties might need to rescale their classical data in order to perform error reconciliation. This is for instance the case of the variable $Y_1$: it is better for Bob to exploit the value of the second mode, $B_1'$, of his two-mode squeezer, which can be obtained by rescaling $Y_1$ by $\sqrt{2(V_B-1)/(V_B+1)}$.
The new covariance matrix is given by:
\begin{align*}
\Gamma_{X_1 X_2 Y'_1 Y_2} = \mathrm{diag}\left(1,1,\sqrt{\frac{2(V_B-1)}{V_B+1}}, 1\right)   \Gamma_{X_1 X_2 Y_1 Y_2} \mathrm{diag}\left(1,1,\sqrt{\frac{2(V_B-1)}{V_B+1}}, 1\right),
\end{align*}
where $\mathrm{diag}(a,\ldots, z)$ is the diagonal matrix with diagonal coefficients $a,\ldots, z$.

The mutual information $I(X_2; (Y_1, Y'_2))$ is given by
\begin{align}
I(X_2; (Y_1, Y_2)) = \log_2 \left[ \frac{\det \Gamma_{X_2} \det \Gamma_{Y'_1 Y_2} }{\det \Gamma_{X_2 Y'_1 Y_2}}\right].
\end{align}

The Holevo information $\chi(X_2;E)$ can be computed as follows:
\begin{align}
\chi(X_2;E) &= S(E) - S(E|X_2) = S(A_1 A_2 B_1 B_2) - S(A_1 B_1 B_2 | X_2)
\end{align}
where we used the fact that Eve's register $E$ can be assumed without loss of generality to purify the systems $A_1 A_2 B_1 B_2$. Moreover, because we assume that Alice performs a heterodyne detection, which projects the state onto a pure coherent state, we also have $S(A_1 B_1 B_2 | X_2) = S(E|X_2)$.

Let us denote by $\nu_1, \nu_2, \nu_3, \nu_4$ the symplectic eigenvalues of $\Gamma_{A_1 A_2 B_1 B_2}$ and by $\tilde{\nu}_1, \tilde{\nu}_2, \tilde{\nu}_3$ the symplectic eigenvalues of $\Gamma_{A_1 B_1 B_2 | X_2}$, which is the covariance matrix of the postmeasurement state after Alice has obtained outcome $X_2$.
Then, the Holevo information is given by
\begin{align}
\chi(X_2;E) &= \sum_{i=1}^4 g(\nu_i) - \sum_{i=1}^3 g(\tilde{\nu}_i),
\end{align}
with the function $g$ defined by
\begin{align}
g(x) := \frac{x+1}{2} \log_2 \left(\frac{x+1}{2}\right) -\frac{x-1}{2} \log_2 \left(\frac{x-1}{2}\right).
\end{align}

Recall that the symplectic eigenvalues of the matrix $\Gamma$ correspond to the standard eigenvalues of $|i \Omega \Gamma|$, or equivalently, to the modulus of the eigenvalues of $i \Omega \Gamma$, where the symplectic form $\Omega$ is given by $\Omega =\bigoplus_{i=1}^d 	\begin{bmatrix} 0 & 1 \\ -1 & 0\\ \end{bmatrix}$, and $d$ is the number of modes.

Finally, the key rate is obtained by optimizing over the choices of the variances $V_A, V_B$, transmittance $T$ and squeezing gain $g$. 

We plot on Fig.~\ref{fig:rate} the asymptotic key rate of the two-way protocol and of the (one-way) no-switching protocol, in the case of a noiseless channel ($\xi=0$), assuming perfect reconciliation efficiency. The key rate of the one-way protocol can be obtained similarly by imposing $V_A=1$, \textit{i.e.}, Alice sends the vacuum to Bob, fixing $T=0$ so that mode $B_2$ contains the vacuum, and getting rid of the final squeezer in Alice's lab (by choosing $g=1$). 
We note that the two-way protocol slightly outperforms the one-way protocol in the regime of ultralow loss ($T$ close to 1).
\begin{figure}[H]
\centering
  \includegraphics[width=0.7\linewidth]{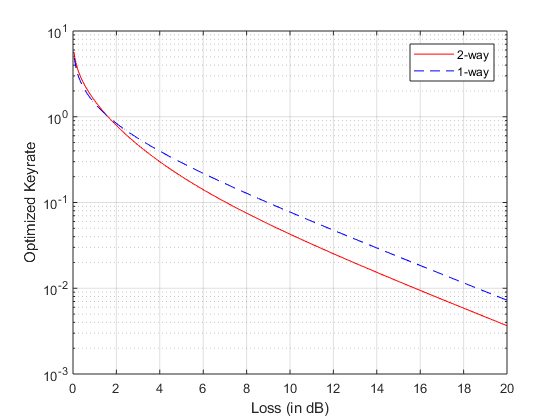}
\caption{Secret key rate of the two-way CV QKD protocol (full line) and of the one-way no-switching protocol (dashed line), assuming a pure-loss channel ($\xi=0$) and perfect reconciliation efficiency ($\beta=1$).}  \label{fig:rate}
\end{figure}

We plot on Fig.~\ref{fig:real} the key rate of the two protocols with a noisy channel ($\xi=0.1$), and realistic reconciliation efficiency, $\beta=0.95$. The advantage of the two-way protocol is clear in this case. 
\begin{figure}[H]
\centering
  \includegraphics[width=0.7\linewidth]{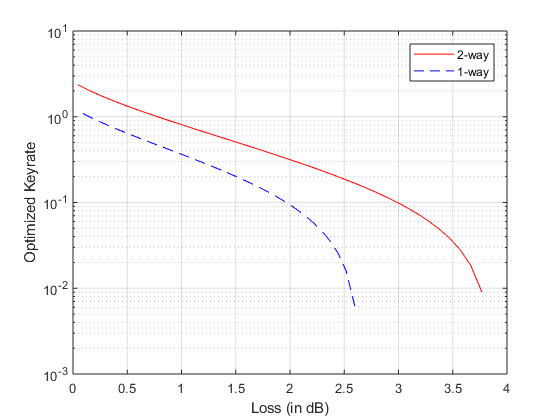}
\caption{Secret key rate of the two-way CV QKD protocol (full line) and of the one-way no-switching protocol (dashed line), assuming a noisy channel ($\xi=0.1$) and imperfect reconciliation efficiency ($\beta=0.95$).}  \label{fig:real}
\end{figure}

We plot on Fig.~\ref{fig:noise} the tolerable excess noise of the two-way and one-way no-switching protocol against the channel transmittance $\tau$, that is the value of $\xi$ for which the key rate becomes 0. The main advantage of the two-way protocol is its much larger tolerance to noise. 
\begin{figure}[H]
\centering
  \includegraphics[width=0.7\linewidth]{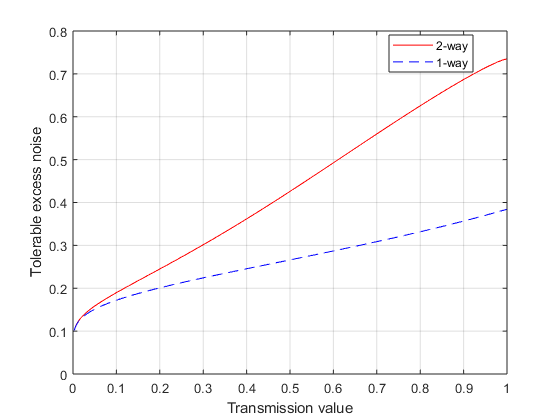}
\caption{Maximum tolerable excess noise $\xi$ for the two-way CV QKD protocol (full line) and the one-way no-switching protocol (dashed line), assuming perfect reconciliation efficiency ($\beta=1$).}  \label{fig:noise}
\end{figure}

\end{widetext}

\end{document}